\documentclass{article}

\usepackage{graphicx}
\usepackage{psfig}
\usepackage{epsfig}
\usepackage[round]{natbib}

\title{
Year-ahead prediction of US landfalling hurricane numbers:
intense hurricanes
}

\author{Stephen Jewson}
\begin{document}

\author{Shree Khare\footnote{\emph{Correspondence address}: Email: \texttt{khare@ucar.edu}}\\and\\
Stephen Jewson\\}

\maketitle

\begin{abstract}
We continue with our program to derive simple practical methods that can be used to
predict the number of US landfalling hurricanes a year in advance. We repeat
an earlier study, but for a slightly different definition
landfalling hurricanes, and for intense hurricanes only.
We find that the averaging lengths needed for optimal predictions of numbers of
intense hurricanes are longer than those needed for optimal predictions of
numbers of hurricanes of all strengths.
\end{abstract}

\section{Introduction}

The reinsurance industry is interested in predictions of hurricane activity on
all timescales, but is \emph{most} interested in predictions that are available
during the reinsurance contract renewal negotiations that typically take place
towards the end of the calendar year. Such predictions can be used directly in
the calculation of annual reinsurance premiums.
Motivated by this, we are attempting to develop methods that give good predictions
of annual hurricane numbers from November of the year prior to the year being predicted.
Since this is a longer lead-time than produced in seasonal forecasts, and since next year's
hurricane count is predicted as soon as we know this year's count, we call this the
\emph{year-ahead} prediction of hurricane numbers.
Our approach is to start with the simplest possible statistical methods, and build
up to more complex methods very gradually. Hopefully in this way we will develop a clear
understanding of the advantages and disadvantages of different methods.

Our first attempt at year-ahead prediction of hurricane numbers was described in~\citet{j81}.
We used simple averages of numbers of historical hurricanes to predict future numbers. The single
parameter in this approach is the length of the average used: we varied this parameter and performed
out-of-sample backtests on the available historical record to see what averaging lengths would have
worked well in the past. We found that shorter windows (from 6 to 28 years) worked better than
longer windows, and this result was shown to be statistically significant. This purely empirical result
corresponds well with physical theories for hurricane numbers that
suggest that there is a multidecadal cycle of hurricane activity~\citep{goldenberg},
most probably governed by sea surface
temperatures as part of a mode of variability known as the Atlantic Multidecadal Oscillation (AMO)~\citep{suttonh}.

The purpose of the current study is twofold.
Firstly, we repeat the previous study, described in~\citet{j81},
but using an alternative, and preferable, definition of landfalling
hurricanes. We will see that the results are more or less the same as before.
Secondly, we repeat the study for \emph{intense} landfalling hurricanes only.
Intense hurricanes cause the most physical damage and so the prediction
of intense hurricanes is of the most interest to reinsurers.

\section{Data}

As in the~\citet{j81} study, the data we use is the HURDAT: the `official' hurricane occurrence data set produced
by the US National Hurricane Centre~\citep{hurdat}. However, as discussed above, we now use a slightly different
definition for landfalling hurricane. The previous study used the \texttt{XING} variable from HURDAT,
while we now use the \texttt{SSS} variable.
The \texttt{XING} variable counts a landfalling hurricane as a storm that is a hurricane at any point in its life, and
that makes landfall. The weakness of this definition is that there are a certain number of hurricanes that
weaken before landfall, and are no longer hurricanes at the moment of landfall. With the \texttt{XING} definition these
are still classified as landfalling hurricanes. However, from the point of view of the on-shore insurance industry,
and others who care about possible on-shore damage caused by hurricanes, such hurricanes are less interesting than
hurricanes that are still hurricanes at the point of landfall.
The \texttt{SSS} variable that we use instead defines `landfalling hurricane' as `a storm that is a hurricane at the
point that it makes landfall'. The definition of `intense landfalling hurricane' that we use
is also derived from the value of the \texttt{SSS} variable.

The number of landfalling hurricanes per year defined using the \texttt{XING} variable is shown in
figure 1 of~\citet{j81}, while the number of landfalling hurricanes per year defined using the
\texttt{SSS} variable is shown in figure~\ref{f01} below. The difference between these two time series
is shown in figure~\ref{f01a}.
In most years this difference is negative or zero, indicating that \texttt{XING} $>$ \texttt{SSS} as expected.
However in 1 year the difference is positive. This would seem to be an inconsistency in the HURDAT database.
However, we use the data as is, as a project is underway elsewhere to reanalyse and correct such inconsistencies.
It seems unlikely that these small inconsistencies will have any influence on our overall conclusions.

The number of intense landfalling hurricanes per year (based on the SSS variable)
is shown in figure~\ref{f01s}.

\section{Method}

For both all hurricanes and intense hurricanes we perform a backtesting study to determine what lengths
of averaging window would have worked well for the year-ahead prediction of hurricane numbers in the past.
The backtesting method used follows the method used in~\citet{j81} exactly.
It consists of:
\begin{enumerate}
    \item Using data from 1940 to 2004, a backtesting comparison to evaluate the use of different
    averaging windows. The skill of each prediction
    is measured using MSE, and the window length that gives the minimum MSE is described as the
    optimal window length.
    \item A statistical test of the optimal window length, based on random reorderings of the observed
    hurricane number time series.
    \item A repeat of the backtesting study using data for 1900 to 2004, as a sensitivity test. This gives
    a second optimal window length.
    \item A repeat of the statistical test, now for the second optimal window length.
    \item A repeat of the backtesting study, as a further sensitivity test,
    now using the 41 data series that start with each year from 1900 to 1940, and that all end with 2004.
\end{enumerate}

\section{Results}

\subsection{All landfalling hurricanes}

The results from the backtesting analysis of all landfalling hurricanes are shown in figures~\ref{f02} to~\ref{f07}.

In figure~\ref{f02} we show the results for the backtesting analysis of data from 1940 to 2004, and we see
 that, as in~\citet{j81}, shorter windows (from 6 to 25 years) work better than very short or very long windows.
The effect is, however, slightly less strong than that seen previously: the minimum in MSE is broader and
the `kink' at 46 years is more pronounced. The lowest value of MSE occurs at 16 years, and the statistical
test gives this a p-value of 3.35\%. Figure~\ref{f04} shows the distribution of optimal window lengths from
the random reordering test.

In figure~\ref{f05} we show the results for the backtesting analysis of data from 1900 to 2004.
In this case, the curve looks very similar to the corresponding curve in~\citet{j81}, with the
minimum lying in a range from around 18 to 40 years. The precise minimum is at 20 years.
The statistical test gives this a p-value of 2\% (40 out of 2000 cases had shorter optimal windows).
Figure~\ref{f06} shows the distribution
of optimal window lengths from the random reordering test, as before.

Finally figure~\ref{f07} gives the distribution of optimal
window lengths derived using start years from 1900 to 1941. The 41 optimal window lengths are distributed
from 6 to 33 years, which is very similar to the distribution shown in~\citet{j81}, figure 7, which was
in the range 6 to 28 years.

Our conclusion at this point is that switching from the \texttt{XING} definition of landfalling to the
\texttt{SSS} definition of landfalling does not make a material difference to the results given in~\citet{j81}:
the number of hurricanes is still best predicted using a short window with a length somewhere
in the range 6 to 33 years.

\subsection{Intense landfalling hurricanes}

We now describe the results from the backtesting analysis of the numbers of \emph{intense} landfalling hurricanes
(where `intense' is defined as Saffir-Simpson category 3-5).
Intense landfalling hurricanes are the most important for many land-dwellers,
because only for these hurricanes are the winds strong enough to cause severe damage to property.

The numbers of intense landfalling hurricanes per year are shown in figure~\ref{f01s}, and the backtesting
results are shown in figures~\ref{f02s} to~\ref{f07s}.
Using data from 1940 to 2004, the best hindcasts were for window lengths from around 15 to 30 years,
with the actual minimum occuring at 16 years (see figure~\ref{f02s}). This has a p-value
of 3.4\% (based on the random reordering results shown in figure~\ref{f04s}, in which 67 out of
2000 cases had shorter optimal windows).
Repeating the analysis for data from 1900 to 2004 gives a slightly
different story, however. The best predictions occurred for window lengths of around 60 years, with the
precise minimum at 56 years. This is noticeably longer than the optimal window lengths in our previous analyses.
However, the distribution of the optimal window length derived from the random reordering is also
noticeably different, with more mass at longer window lengths (see figure~\ref{f06s})
and the value of 56 years still has a p-value of only 8.3\%.
In figure~\ref{f07s} we show the optimal window lengths obtained using data periods starting from
1900 to 1941, and we see values from 16 to 56.

From these results we conclude that the predictability properties of intense hurricanes are different from those
of all hurricanes: longer averaging windows are needed.

\subsection{Sources of forecast error variance}

Figures~\ref{f02}, \ref{f05}, \ref{f02s} and \ref{f05s} all show the decomposition of the MSE into
bias (dotted line) and variance (solid line) terms.
For short windows we see that the variance term completely dominates.
This term can be decomposed further, into `internal variability' and `sampling error' terms.
We show this decomposition for the four variance curves in these four figures in figure~\ref{f08}.
What we see is that in all cases the reason that very short windows give poor predictions is mostly because
of the high level of sampling error, as we might expect.

\section{Conclusions}

We are interested in predicting the number of landfalling hurricanes
and the number of intense landfalling hurricanes, one year in advance.
We are investigating using very simple time averages of historical hurricane
numbers to make these predictions. Specifically, we have performed two studies:
\begin{itemize}
    \item We have repeated the analysis of~\citet{j81}, but for a more appropriate
    and more standard definition of `landfalling hurricane'.
    \item We have repeated the analysis of~\citet{j81}, but for intense landfalling
    hurricanes only.
\end{itemize}

The results are as follows.
For all landfalling hurricanes, we find more or less the same results as we found in~\citet{j81}, which are
that:
\begin{itemize}
    \item Short averaging windows, in the range from 6 years to 33 years, would have given the best predictions
    \item But there is a lot of uncertainty around any point estimate of the best window length
\end{itemize}

For intense landfalling hurricanes, the results are interestingly different from the results for all hurricanes.
Again we found that short windows give the best predictions. But now the window lengths are longer: from 16
to 56 years.
This suggests that the nature of the predictability of intense hurricanes is different from that of all
hurricanes.
It seems possible that there may be a purely statistical explanation
for this: intense hurricanes are less common, and as a result
the intense hurricane time series has a lower signal to noise
ratio than the time series for all hurricanes.
However whether this can really explain the differences that we see
needs further, and more detailed, investigation.

Where does this leave us with respect to making practical forecasts of hurricane numbers (intense or otherwise)
a year in advance? The biggest hurdle in making such predictions seems to be the problem of which
window length to choose from within the wide ranges that emerge from our analysis. For all hurricanes, we could choose
any number from 6 to 33 years. For intense hurricanes, we could choose any number from 16 to 56 years.
These different choices would give very different results. How can we reduce this arbitrariness in a
non-arbitrary way? Currently the most attractive answer seems to be to use weighting of each of these
forecasts (or of forecasts using all possible window lengths), where the weighting is based on the likelihood
scores achieved in the backtesting experiment. Testing this is one of our future priorities.

\section{Legal statement}

SJ was employed by RMS at the time that this article was written.

However, neither the research behind this article nor the writing
of this article were in the course of his employment, (where 'in
the course of their employment' is within the meaning of the
Copyright, Designs and Patents Act 1988, Section 11), nor were
they in the course of his normal duties, or in the course of
duties falling outside his normal duties but specifically assigned
to him (where 'in the course of his normal duties' and 'in the
course of duties falling outside his normal duties' are within the
meanings of the Patents Act 1977, Section 39). Furthermore the
article does not contain any proprietary information or trade
secrets of RMS. As a result, the authors are the owners of all the
intellectual property rights (including, but not limited to,
copyright, moral rights, design rights and rights to inventions)
associated with and arising from this article. The authors reserve
all these rights. No-one may reproduce, store or transmit, in any
form or by any means, any part of this article without the
authors' prior written permission. The moral rights of the authors
have been asserted.

The contents of this article reflect the authors' personal
opinions at the point in time at which this article was submitted
for publication. However, by the very nature of ongoing research,
they do not necessarily reflect the authors' current opinions. In
addition, they do not necessarily reflect the opinions of the
authors' employers.

\bibliography{../bib/jewson}

\newpage
\begin{figure}[!htb]
  \begin{center}
    \scalebox{0.8}{\includegraphics{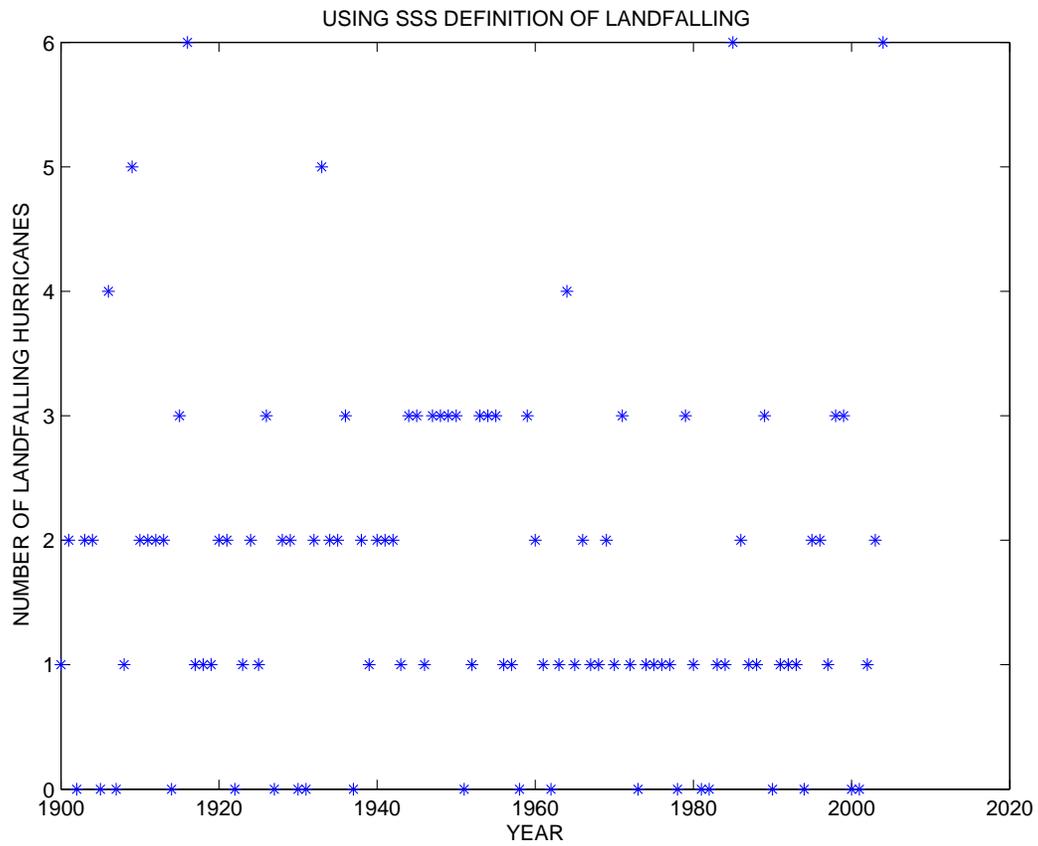}}
  \end{center}
  \caption{
The observed number of US landfalling hurricanes for each year since 1900, using the
SSS definition of landfalling from the HURDAT database.
          }
  \label{f01}
\end{figure}

\newpage
\begin{figure}[!htb]
  \begin{center}
    \scalebox{0.8}{\includegraphics{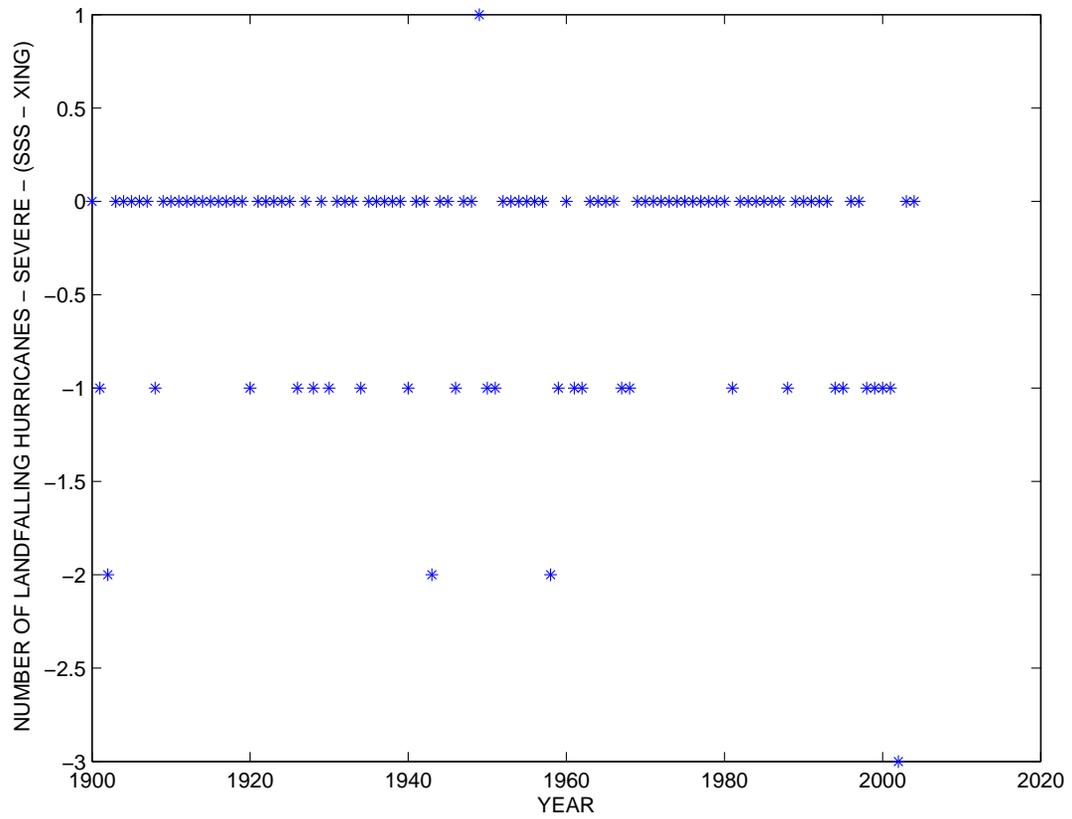}}
  \end{center}
  \caption{
The difference between the number of US landfalling hurricanes calculated using the SSS
definition and the XING definition.
          }
  \label{f01a}
\end{figure}

\newpage
\begin{figure}[!htb]
  \begin{center}
    \scalebox{0.8}{\includegraphics{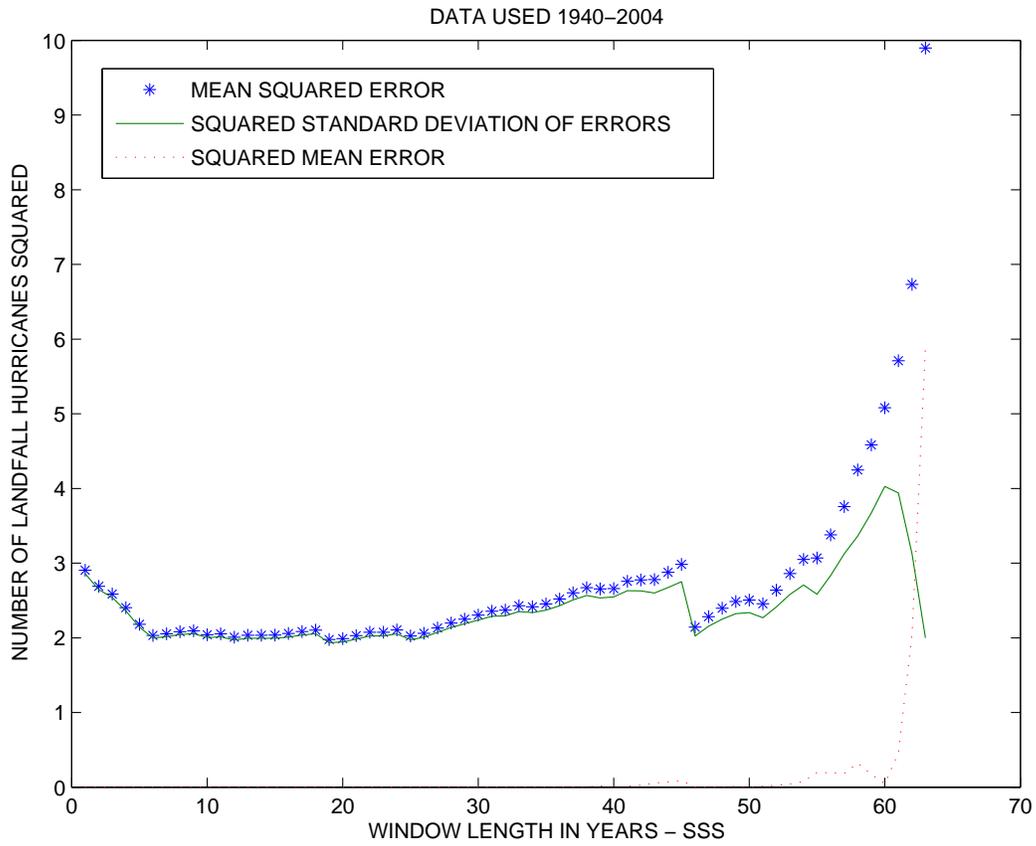}}
  \end{center}
  \caption{
The results from a backtesting study of the ability of averages of length $n$ years to
predict the time series of numbers of US landfalling hurricanes.
          }
  \label{f02}
\end{figure}

\newpage
\begin{figure}[!htb]
  \begin{center}
    \scalebox{0.8}{\includegraphics{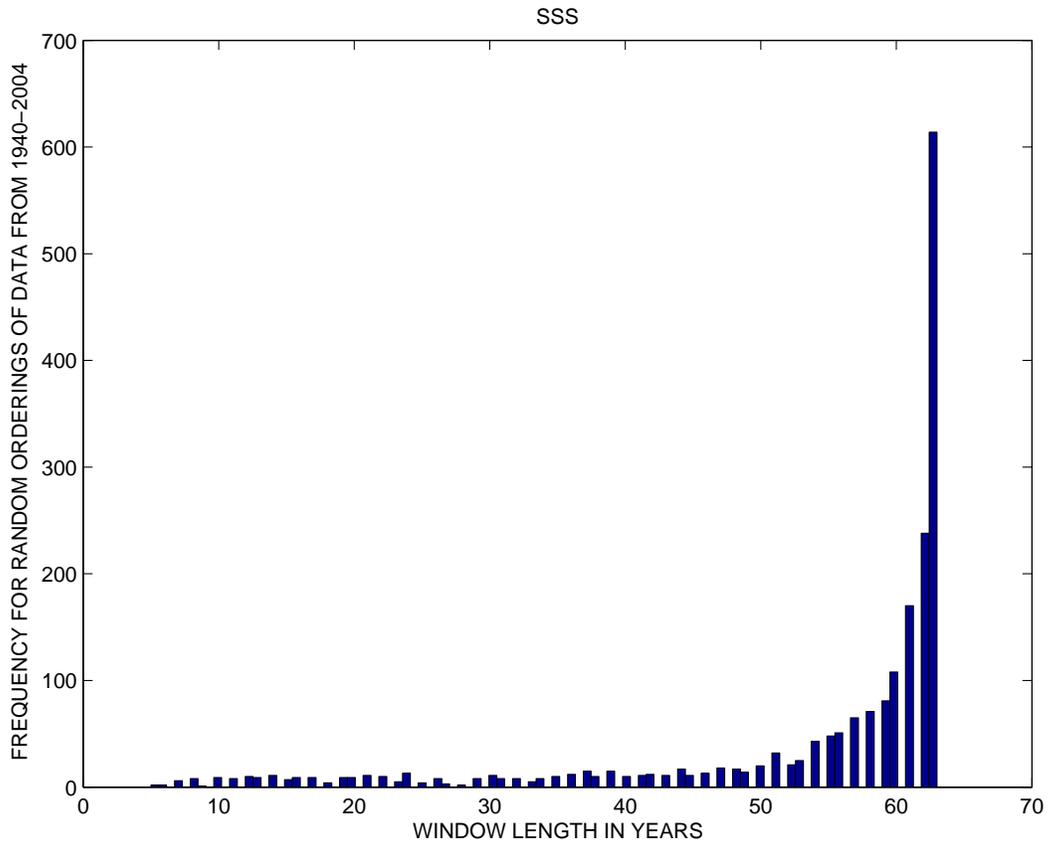}}
  \end{center}
  \caption{
The results from a statistical test of the minimum in figure~\ref{f02}.
In 2000 random reorderings of the hurricane number time series, 67
fall below 16 years, giving a p-value of 3.35\%.
          }
  \label{f04}
\end{figure}

\newpage
\begin{figure}[!htb]
  \begin{center}
    \scalebox{0.8}{\includegraphics{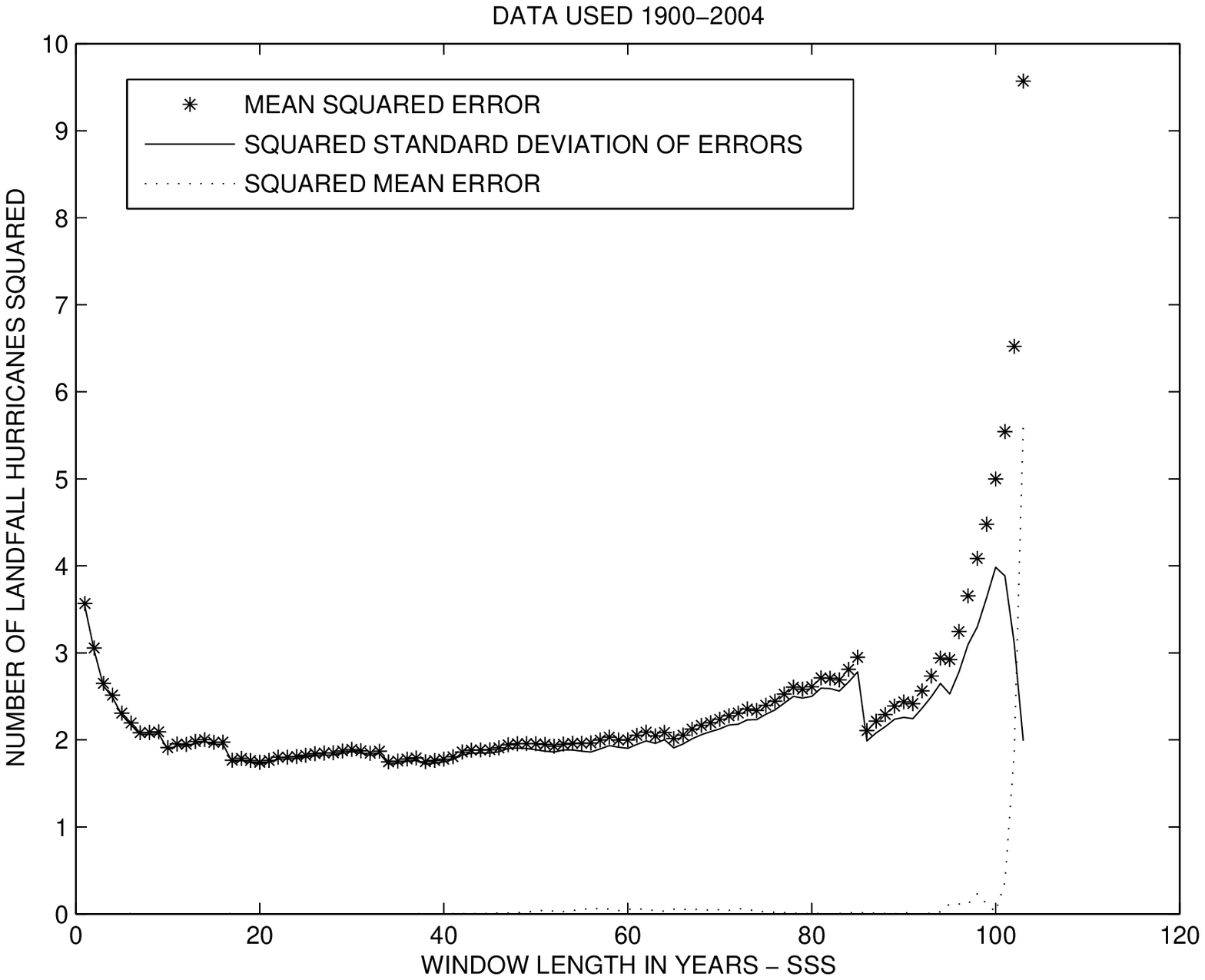}}
  \end{center}
  \caption{
As figure~\ref{f02}, but now using data from 1900 to 2004.
          }
  \label{f05}
\end{figure}

\newpage
\begin{figure}[!htb]
  \begin{center}
    \scalebox{0.8}{\includegraphics{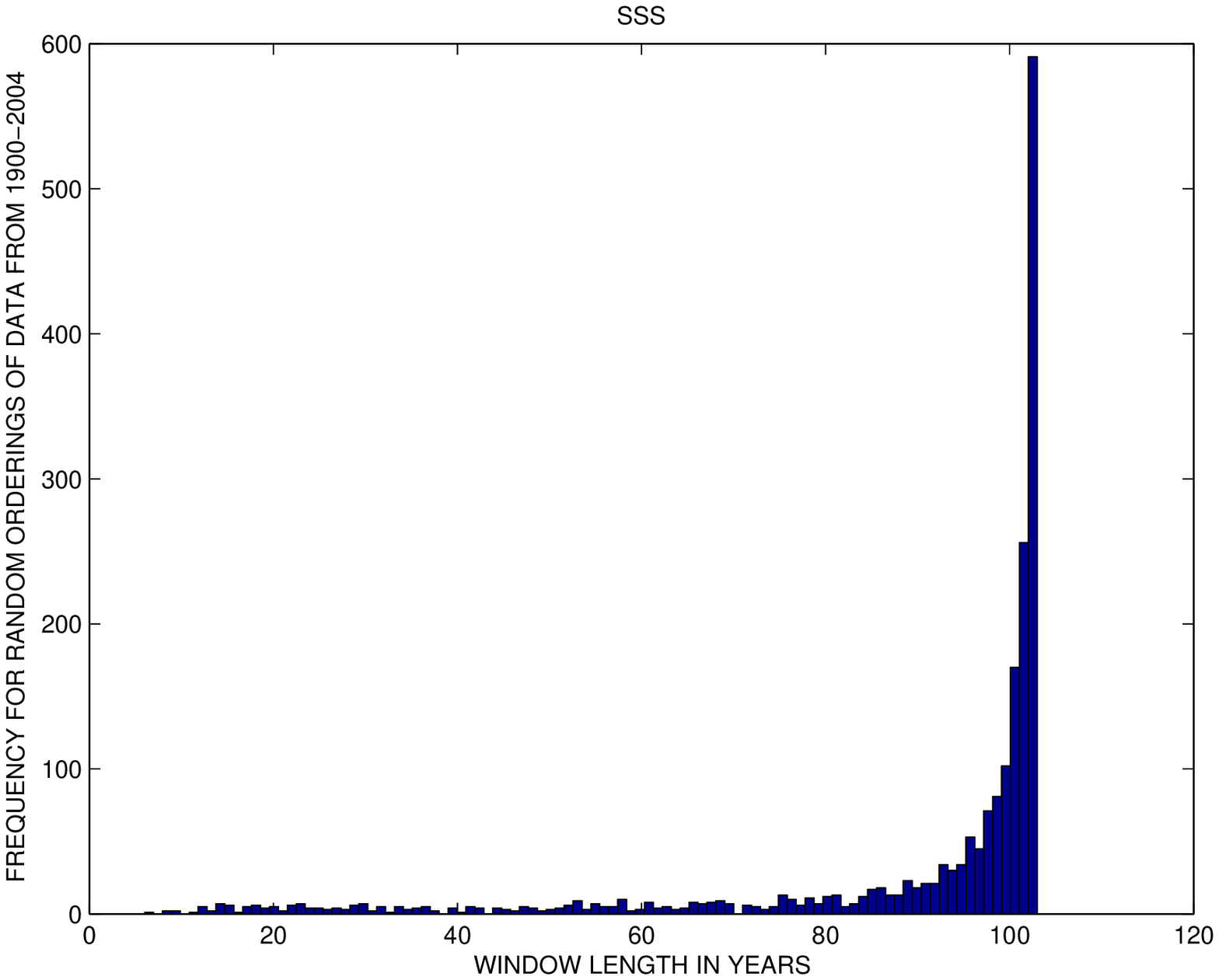}}
  \end{center}
  \caption{
As figure~\ref{f04}, but now using data from 1900 to 2004.
          }
  \label{f06}
\end{figure}

\newpage
\begin{figure}[!htb]
  \begin{center}
    \scalebox{0.8}{\includegraphics{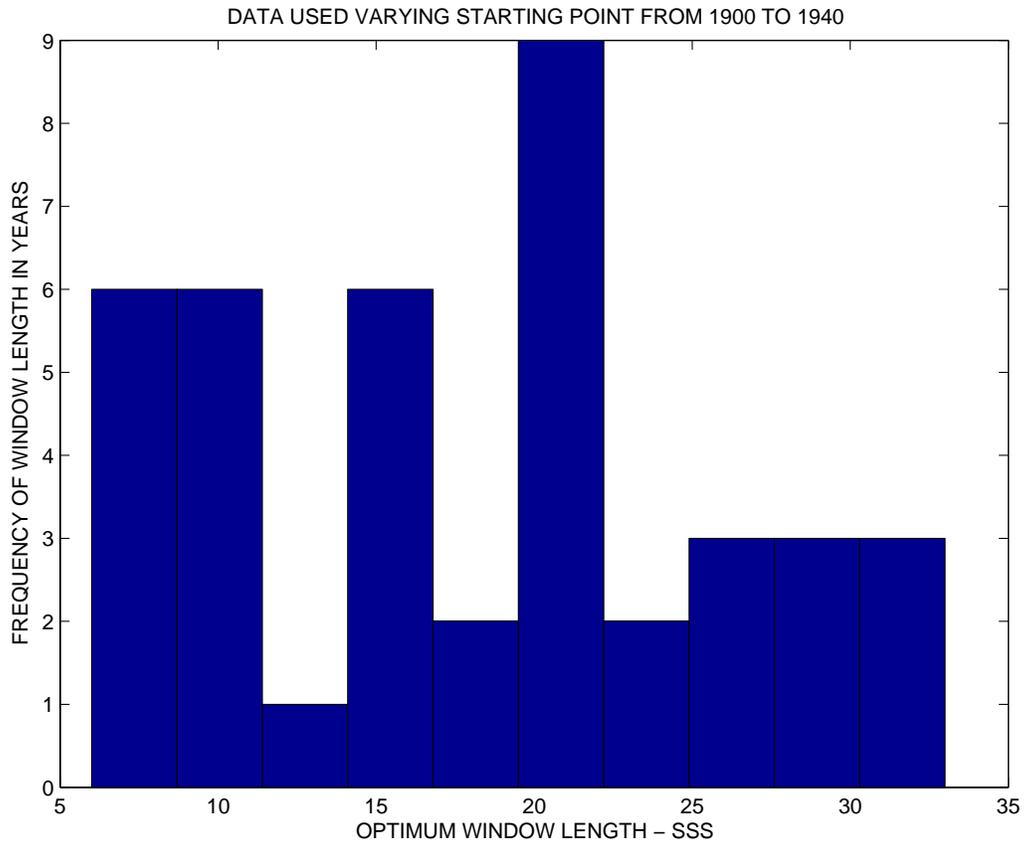}}
  \end{center}
  \caption{
The distribution of optimal window lengths from our backtesting
comparison of methods for prediction of the number of landfalling
hurricanes. The shortest optimal window length is 6 years and the
longest is 33 years.
          }
  \label{f07}
\end{figure}

\newpage
\begin{figure}[!htb]
  \begin{center}
    \scalebox{0.8}{\includegraphics{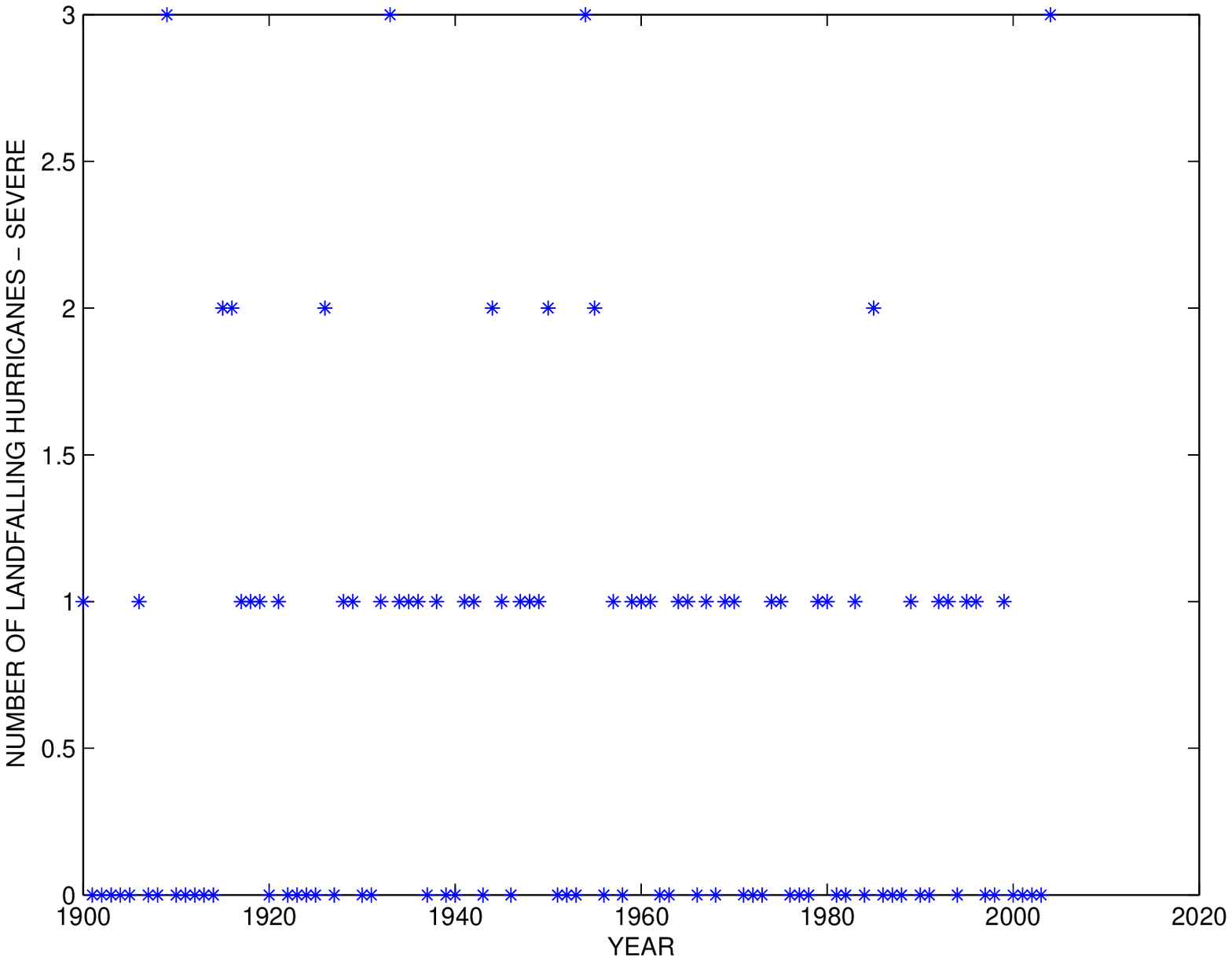}}
  \end{center}
  \caption{
The observed number of \emph{intense} US landfalling hurricanes for each year since 1900, using the
SSS definition of landfalling from the HURDAT database.
          }
  \label{f01s}
\end{figure}

\newpage
\begin{figure}[!htb]
  \begin{center}
    \scalebox{0.8}{\includegraphics{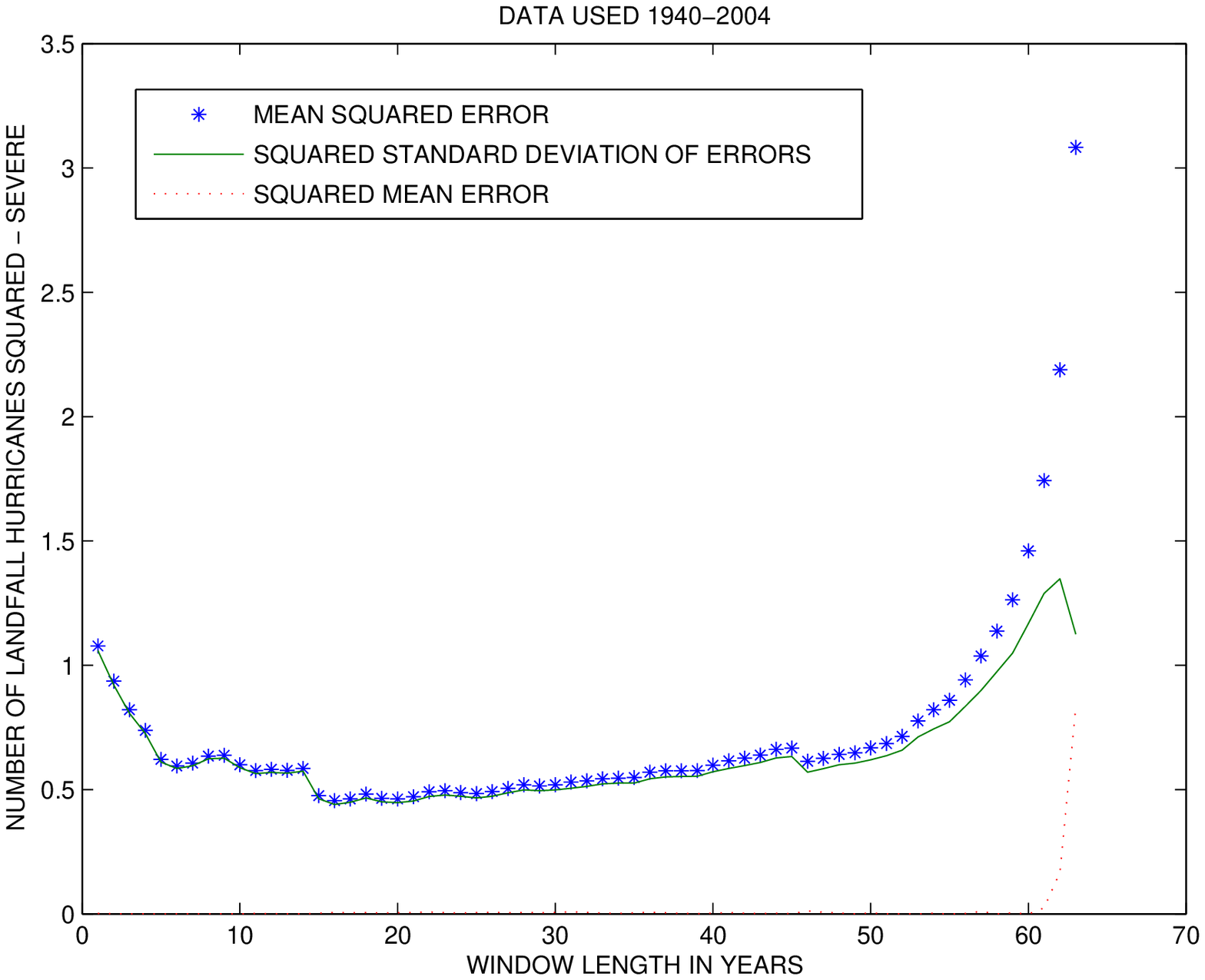}}
  \end{center}
  \caption{
As figure~\ref{f02}, but now for intense hurricanes.
          }
  \label{f02s}
\end{figure}

\newpage
\begin{figure}[!htb]
  \begin{center}
    \scalebox{0.8}{\includegraphics{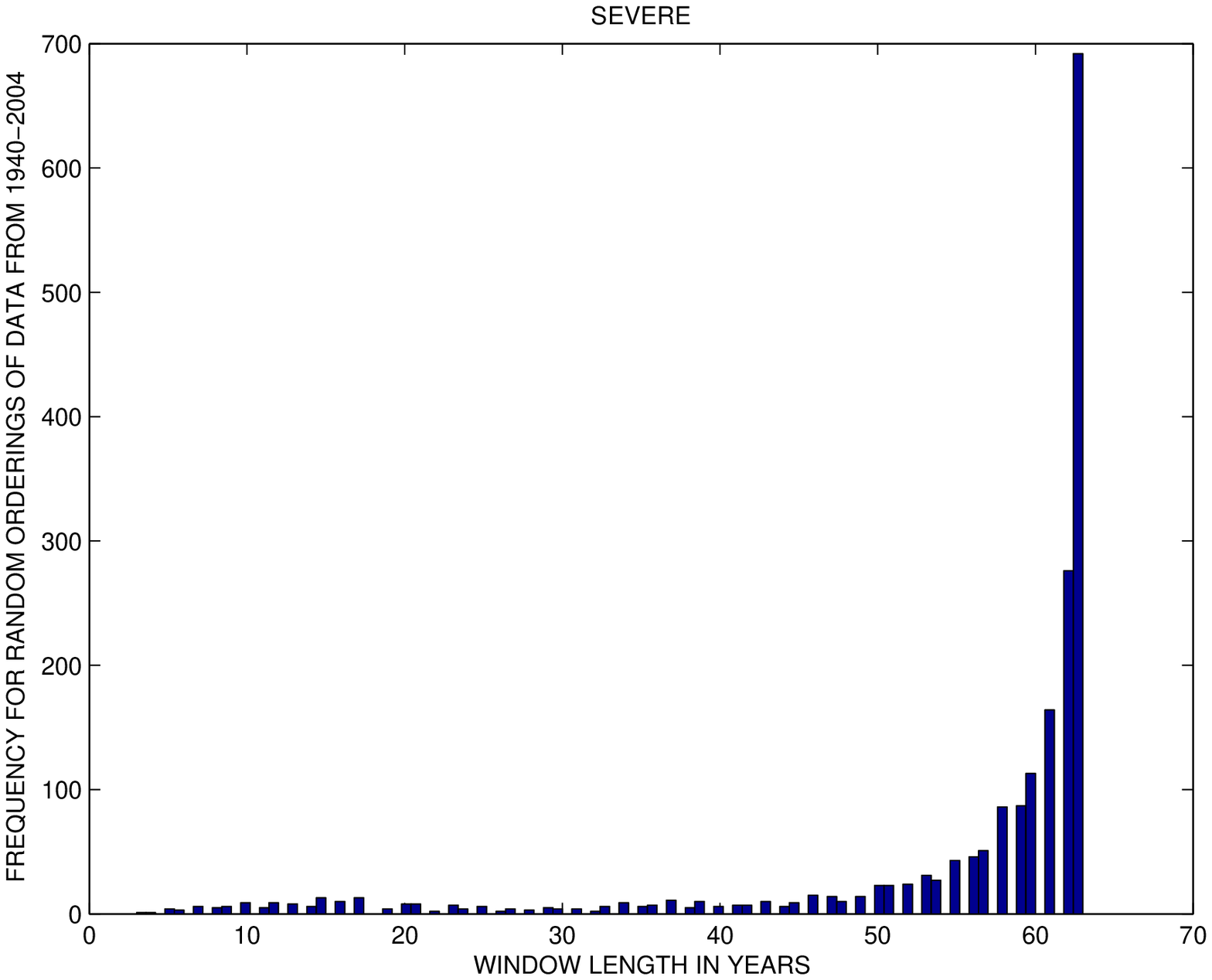}}
  \end{center}
  \caption{
As figure~\ref{f04}, but now for intense hurricanes.
          }
  \label{f04s}
\end{figure}

\newpage
\begin{figure}[!htb]
  \begin{center}
    \scalebox{0.8}{\includegraphics{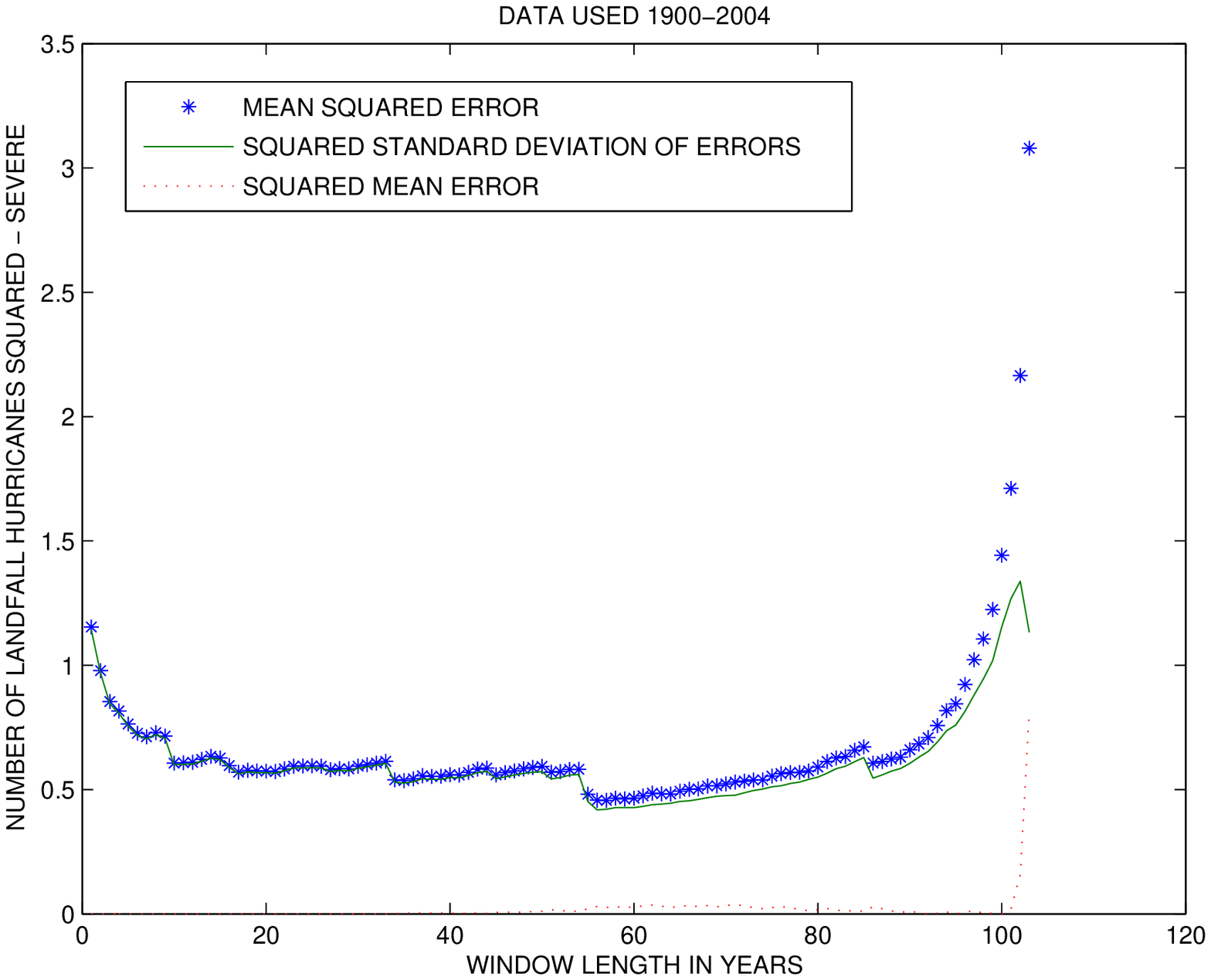}}
  \end{center}
  \caption{
As figure~\ref{f05}, but now for intense hurricanes.
          }
  \label{f05s}
\end{figure}

\newpage
\begin{figure}[!htb]
  \begin{center}
    \scalebox{0.8}{\includegraphics{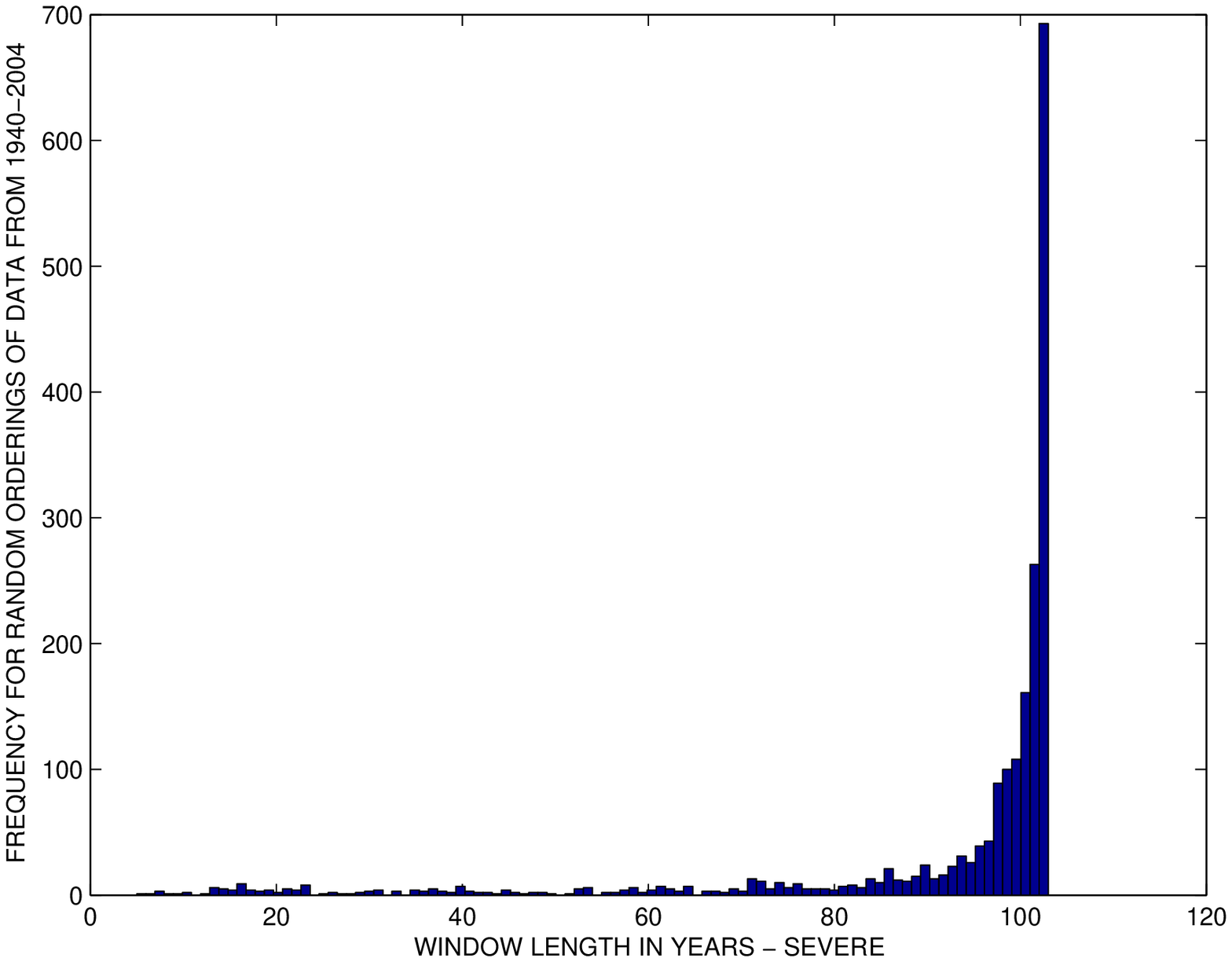}}
  \end{center}
  \caption{
As figure~\ref{f06}, but now for intense hurricanes.
          }
  \label{f06s}
\end{figure}

\newpage
\begin{figure}[!htb]
  \begin{center}
    \scalebox{0.8}{\includegraphics{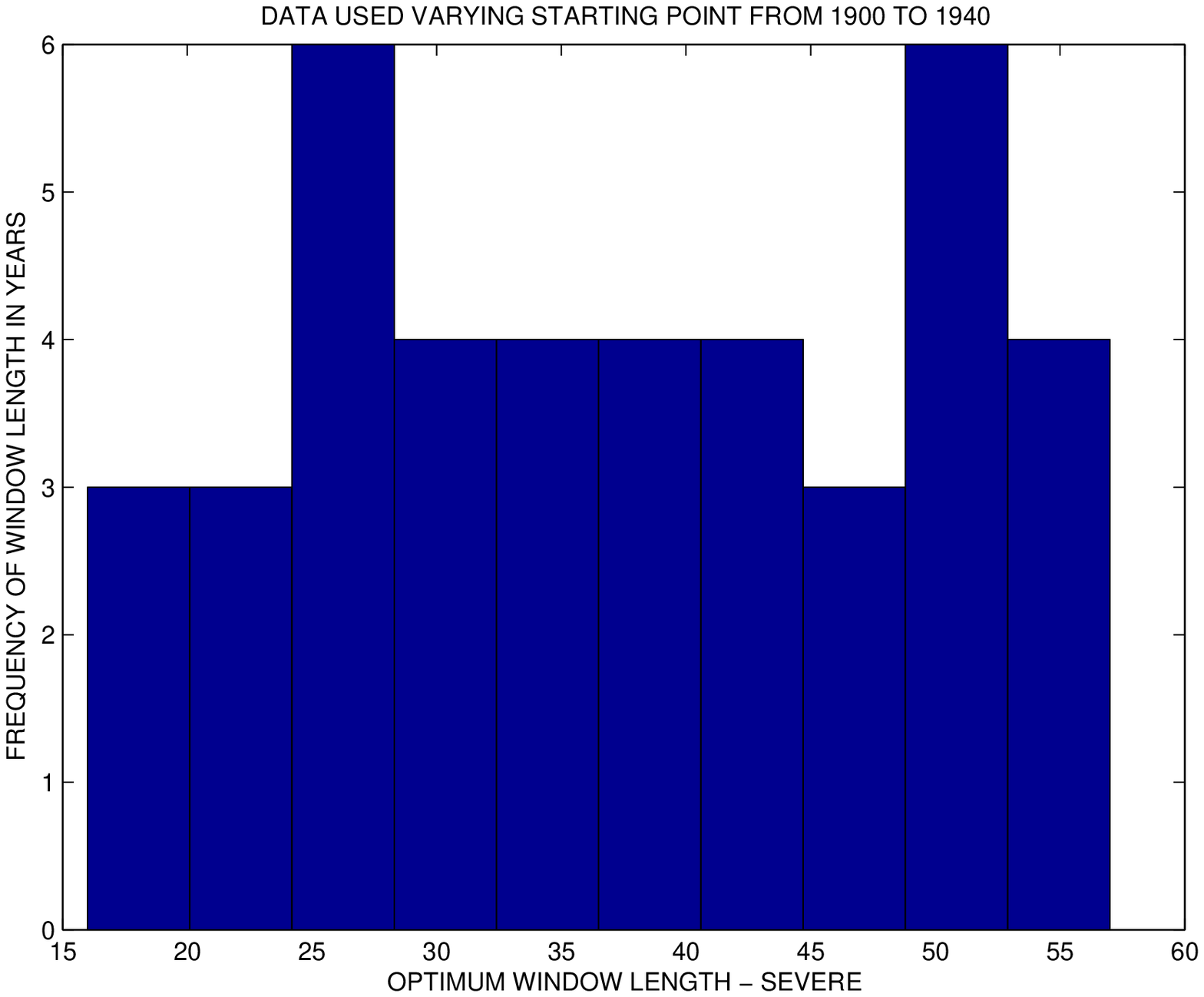}}
  \end{center}
  \caption{
As figure~\ref{f07}, but now for intense hurricanes.          }
  \label{f07s}
\end{figure}

\newpage
\begin{figure}[!htb]
  \begin{center}
    \scalebox{0.8}{\includegraphics{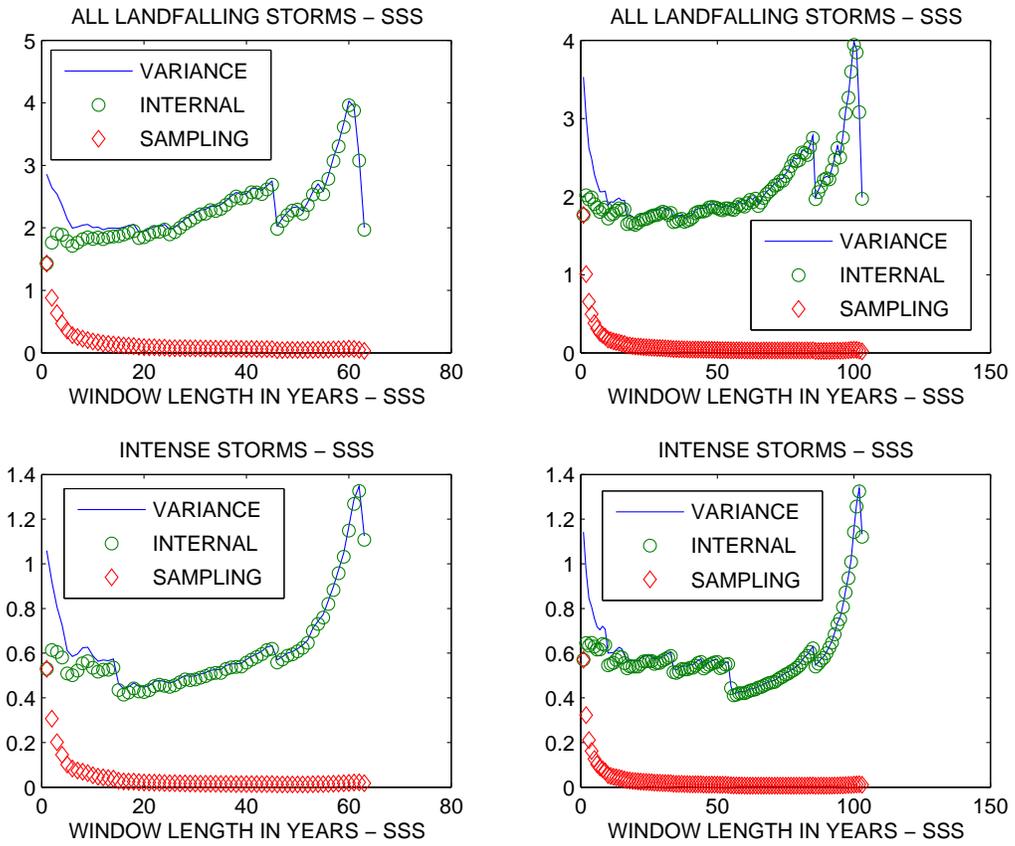}}
  \end{center}
  \caption{
The decomposition of the variance curves shown in figures~\ref{f02},
\ref{f05}, \ref{f02s} and \ref{f05s} into internal variability variance
and sampling error variance.
          }
  \label{f08}
\end{figure}

\end{document}